\documentclass[sigconf]{acmart}

\usepackage{pifont}
\usepackage{amsmath,amsfonts,bm}
\usepackage{algorithm}
\usepackage{algorithmic}

\usepackage{multirow}
\usepackage{cleveref}
\usepackage{upgreek}
\usepackage{threeparttable}
\usepackage{balance}
\crefformat{section}{\S#2#1#3} 
\crefformat{subsection}{\S#2#1#3}
\crefformat{subsubsection}{\S#2#1#3}
\theoremstyle{definition}
\newtheorem{definition}{\textbf{Definition}}

\AtBeginDocument{%
  \providecommand\BibTeX{{%
    \normalfont B\kern-0.5em{\scshape i\kern-0.25em b}\kern-0.8em\TeX}}}

\acmYear{2020}\copyrightyear{2020}
\setcopyright{acmcopyright}
\acmConference[BuildSys '20]{The 7th ACM International Conference on Systems for Energy-Efficient Buildings, Cities, and Transportation}{November 18--20, 2020}{Virtual Event, Japan}
\acmBooktitle{The 7th ACM International Conference on Systems for Energy-Efficient Buildings, Cities, and Transportation (BuildSys '20), November 18--20, 2020, Virtual Event, Japan}
\acmPrice{15.00}
\acmDOI{10.1145/3408308.3427982}
\acmISBN{978-1-4503-8061-4/20/11}

\begin{document}

\title{Kalibre: Knowledge-based Neural Surrogate Model Calibration for Data Center Digital Twins}

\settopmatter{authorsperrow=4}

\author{Ruihang Wang}
\affiliation{%
  \institution{Nanyang Technological \\ 
  	University, Singapore}
}

\author{Xin Zhou}
\affiliation{%
	\institution{Nanyang Technological \\
		University, Singapore}
}

\author{Linsen Dong}
\affiliation{%
	\institution{Nanyang Technological \\
		University, Singapore}
}

\author{Yonggang Wen}
\affiliation{%
	\institution{Nanyang Technological \\
		University, Singapore}
}

\author{Rui Tan}
\affiliation{%
	\institution{Nanyang Technological \\
		University, Singapore}
}

\author{Li Chen}
\affiliation{%
	\institution{Alibaba Inc.}
	\country{China}
}

\author{Guan Wang}
\affiliation{%
	\institution{Alibaba Inc.}
	\country{China}
}

\author{Feng Zeng}
\affiliation{%
	\institution{Alibaba Inc.}
	\country{China}
}

\renewcommand{\shortauthors}{R. Wang, et al.}

\begin{abstract}
	Computational fluid dynamics (CFD) model has been widely used for prototyping data centers. Evolving it to high-fidelity {\em digital twin} is desirable for the management and operations of large-scale data centers. Manually calibrating CFD model parameters to achieve twin-class fidelity by specially trained domain expert is tedious and labor-intensive. To reduce manual efforts, existing automatic calibration approaches developed for various computational models apply heuristics to search model configurations within an empirically defined parameter bound. However, in the context of CFD, each search step requires long-lasting CFD model's iterated solving, rendering these approaches impractical with increased model complexity. This paper presents Kalibre, a knowledge-based neural surrogate approach that performs CFD model calibration by iterating four key steps of i) training a neural surrogate model based on CFD-generated data, ii) finding the optimal parameters at the moment through neural surrogate retraining based on sensor-measured data, iii) configuring the found parameters back to the CFD model, and iv) validating the CFD model using  sensor-measured data as the ground truth. Thus, the parameter search is offloaded to the neural surrogate which is ultra-faster than CFD model's iterated solving. To speed up the convergence of Kalibre, we integrate prior knowledge of the twinned data center's thermophysics into the neural surrogate design to improve its learning efficiency. With about five hours computation on a 32-core processor, Kalibre achieves mean absolute errors (MAEs) of 0.81\textdegree{}C and 0.75\textdegree{}C in calibrating two CFD models for two production data halls hosting thousands of servers each while requires fewer CFD solving processes than existing baseline approaches.
\end{abstract}


\maketitle

\section{Introduction}
\label{sec1}
To meet the ever increasing cloud computing and storage demands, the scales of modern data centers have been continuously growing. According to a white paper from Cisco \cite{Cisco}, the number of hyperscale data centers will double from 338 at the end of 2016 to 628 by 2021. The data centers' increases in size and complexity bring substantial challenges to the effective and efficient management of their supporting infrastructures for avoiding operational risks and reducing energy costs. Currently, data center infrastructure management (DCIM) system is a common tool that visualizes and monitors the infrastructure status based on the measurements collected from deployed sensors~\cite{wiboonrat2014data}. DCIM provides the operator with useful and important information for proper responses in case of abnormalities and failures. However, with the increases in system scale and complexity, it is important to extend DCIM to have accurate prediction capabilities. With such, the operator can perform various what-if analyses, such as whether the increase of certain temperature setpoints can improve the energy efficiency without causing server overheating.

We consider {\em digital twin} for the desired capability extension. Digital twin is a collection of integrated multi-physics, multi-scale, and probabilistic modeling and simulation techniques for as-built systems~\cite{shafto2010draft}. It aims to pursue high modeling accuracy for complex systems based on data from various sources, including sensors, prior models, and domain knowledge. The concept was early applied in the aerospace industry and is now attracting interest in smart manufacturing~\cite{qi2018digital}, cyber-physical systems~\cite{alam2017c2ps}, and smart city creation~\cite{mohammadi2017smart}. To build digital twins for data centers, various elementary techniques from multiple disciplines have existed to model the cyber-physical processes from the building level to the chip level. In particular, the computational fluid dynamics (CFD) modeling is a primary technique to characterize the thermodynamics in data centers~\cite{radmehr2013cfd}. It has been adopted in the offline analysis for energy cost reduction and risk management~\cite{ran2019deepee}. However, the accuracy of the CFD models in general does not reach the digital twin class for online operations. This is because the assumptions or simplifications made in the prototyping phase may lead to result distortions.

To evolve a CFD model into its digital twin form, it is important to instrument the model with sufficiently complete configuration of the physical infrastructure. A model with incomplete configuration may diverge from the ground truth. For example, as reported in~\cite{watson1997design, singh2010cfd}, a manually constructed CFD model can yield temperature prediction errors up to 5\textdegree{}C. Unfortunately, obtaining the complete system configuration often faces substantial challenges due to 1) the large number of parameters in the configuration and 2) the labor-intensive and error-prone manual calibration process for these parameters. For instance, each server in a data center may have its own characteristics of the passing-through air flow rate due to its internal fan control logic. However, such information is often not available in the server hardware's specification and can only be empirically estimated or manually collected via {\em in situ} measurement.

To achieve twin-class accuracy, automatic calibration of the difficult-to-obtain system configuration parameters will be necessary. However, this turns out to be a challenging task, due primarily to the ultra-high computation overhead of executing the CFD model to assess any candidate parameter configuration. The existing heuristic approaches (e.g., evolution strategies, genetic algorithms, simulated annealing, etc.) can be applied for automatic CFD calibration~\cite{rechenberg1973evolutionsstrategie,holland1992genetic}. These approaches in general require many search iterations, e.g., hundreds (cf.~\cref{sec5}), to find good settings for the system configuration parameters. In each iteration, a CFD model solving is performed with the candidate configuration. When the CFD is built for a large-scale data hall with fine meshing granularity, the iterative search process incurs unacceptable computation times, since a single CFD model solving is already an hours-long or even days-long computing process of solving the Navier-Stokes formulation~\cite{anderson1995computational} using the finite element method. As such, the existing search-based approaches scale poorly with the size and granularity of the CFD model.

\begin{figure}[t]
	\centering
	\includegraphics[width=1\columnwidth]{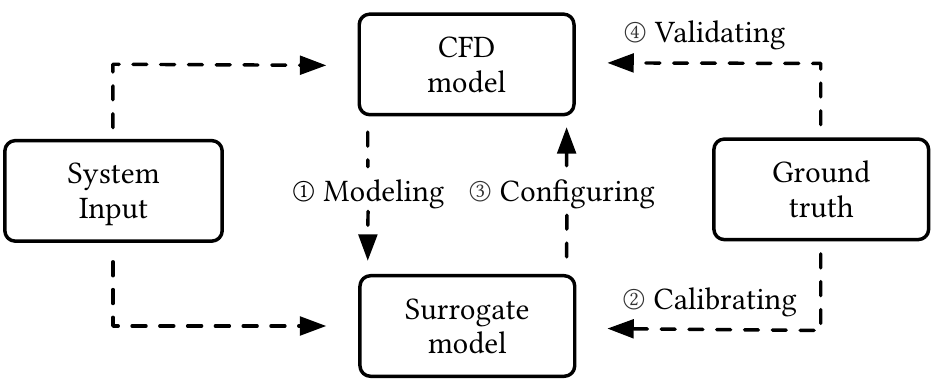}
	\caption{Kalibre operates by iterating \ding{172} training a CFD surrogate model, \ding{173} searching optimal parameters through surrogate retraining, \ding{174} configuring the found parameters back to the CFD model, \ding{175} validating the new configuration.}
	\label{fig1}
\end{figure}

To advance automatic calibration, we propose Kalibre, a neural surrogate-assisted approach to calibrate data center CFD models with increasing scales and complexities. Kalibre avoids directly solving the CFD model for configuration search with the help of a trainable neural net. Fig.~\ref{fig1} illustrates Kalibre's workflow, in which the surrogate model iteratively updates the system configuration to minimize the CFD model's prediction errors by four key steps. \ding{172} The "coarse" surrogate is trained to align with the "fine" CFD model in the current system state locality by updating its internal weights based on CFD-generated data. \ding{173} The trained surrogate is re-optimized by updating the system configuration, which is also a part of trainable variables of the neural net, to maximize the consistency between the surrogate's predictions and the ground-truth sensor measurements. \ding{174} The updated system configuration is set back to the CFD model for refining. \ding{175} The ground-truth sensor measurements are used to validate the refined CFD model. Therefore, Kalibre offloads the fine-grained parameter configuration search to the surrogate. Vis-{\`a}-vis the existing heuristic approaches that solve the CFD model every configuration search step, Kalibre solves the CFD model much less frequently for merely providing feedback to the surrogate.

The implementation of Kalibre faces two challenges. First, the design of the surrogate to capture the complex thermophysics encompassed in the CFD model is challenging.
Second, the training data for the neural surrogate is limited since generating such data using the CFD model is compute-intensive. Piecemeal solutions to address the above two challenges separately are contradictory, i.e., a deeper neural surrogate to well capture the complex thermodynamics requires more CFD-generated training data.    
To address the challenges, we design a neural surrogate architecture that integrates the prior knowledge of the thermal relations among a number of key variables in the twinned data hall.
Compared with the vanilla neural net that approximates the CFD model as a black box, the introduction of the prior knowledge regulates the number of trainable variables and significantly improves the learning efficiency on small data.

We implement Kalibre and apply it to calibrate the CFD models of two production data halls sized hundreds of square meters that host thousands of servers, respectively. The CFD models calibrated by Kalibre achieve mean absolute errors (MAEs) of 0.81\textdegree{}C and 0.75\textdegree{}C in predicting the temperatures at tens of cold/hot aisle positions in each hall, respectively. The calibration process takes about 5 hours on a 32-core virtual machine in the cloud. In contrast, the heuristic configuration search and the vanilla neural net-based surrogate approach achieve MAEs of around 1.5$\sim$4\textdegree{}C with the same computation time for calibration as Kalibre. We also invite a domain expert to manually fine-calibrate the two CFD models, yielding MAEs of 1.32\textdegree{}C and 1.1\textdegree{}C, which are higher than Kalibre's MAEs by 63\% and 46\%, respectively. The absolute MAE reductions of 0.51\textdegree{}C and 0.35\textdegree{}C achieved by Kalibre in comparison with the expert's manual calibration are significant in CFD modeling, due to the sharply increased difficulty in improving accuracy when the errors are already low (i.e., at around 1\textdegree{}C). The evaluation shows the effectiveness of Kalibre in automatically calibrating CFD models toward their digital twin forms.

{\bf Roadmap:} The rest of this paper is structured as follows. \cref{sec2} reviews related work. \cref{sec3} formulates problem and overviews our approach. \cref{sec4} and \cref{sec5} design and evaluate Kalibre, respectively. \cref{sec:discuss} discusses several issues and concludes this paper.

\section{Related Work}
\label{sec2}
This section reviews the relevant studies in data center modeling, surrogate-assisted optimization and knowledge-based neural nets.

{\bf Data center modeling:}
A variety of modeling techniques have been proposed for thermal management in data centers. They can be broadly categorized into law-based, data-driven, and hybrid models. The CFD models are representative law-based models, in that they capture the thermodynamic laws followed by the physical processes~\cite{radmehr2013cfd, singh2010cfd, ran2019deepee}. However, the CFD models are computationally expensive due to their internal recursive execution. As the mesh complexity increased for large-scale data centers, the CFD models solving times may increase from hours to days, introducing significant challenges for model calibration.
An alternative is to use black-box data-driven models to learn a thermal map in the data center. For example, the Weatherman system \cite{moore2006weatherman} predicts the steady-state temperatures of certain server blocks using a neural net consisting of two hidden layers. In \cite{yi2019toward}, a long short-term memory (LSTM) network is designed for predicting server CPU temperature. Although these data-driven models are fast and suitable for real-time use, they often perform poorly in the cases that are not covered by the training data. For instance, these models cannot well capture the thermal processes in case of cooling system failures, because the training data for such failure scenarios is generally lacking.
Hybrid methods of combining CFD models and data-driven models have also been proposed. For instance, in~\cite{van2019control}, a psychrometric model is jointly used with three multilayer perceptrons to predict steady system state. In \cite{chen2012high}, the actual dataset is augmented with CFD-generated data for rare scenarios; the augmented dataset is used to train a linear regression model for temperature prediction. To ensure fidelity, the CFD model used in~\cite{chen2012high} is manually fine-calibrated by a human expert. As such, the approach is only evaluated on a small-size testbed and cannot scale well with the data center size.

{\bf Surrogate-assisted optimization:}
Surrogate-assisted optimization~\cite{koziel2012surrogate} speeds up the parametric optimization of those compute-intensive and non-differentiable models. It builds a lightweight surrogate of the original model and then uses the surrogate to guide the parameter search. This technique has been applied to building energy~\cite{nagpal2019methodology}, hydrological~\cite{asher2015review}, and aerodynamic~\cite{zhonghua2019efficient} model optimization. The surrogate design is application-specific. For example, low-fidelity law-based surrogate model is built for full-fledged models in microwave engineering by aggressive space mapping~\cite{bandler1995electromagnetic}. Data-driven surrogate based on neural net is designed for high-dimensional and nonlinear coplanar waveguide model~\cite{watson1997design}. Response surface methodology based on radial basis function is studied for CFD model~\cite{phan2019cfd}. Among these studies, data-driven surrogates exhibit advantage in fast forwarding. However, the design of surrogate-assisted optimization faces a general challenge in well balancing the surrogate fidelity and the computation overhead of generating training data for surrogate via executing the original compute-intensive models. An effective approach to addressing the challenge is to improve the learning efficiency of the data-driven surrogate via its architectural design. Unfortunately, few studies are dedicated to pursuing surrogate's learning efficiency in the context of data center CFD.

{\bf Knowledge-based neural nets:} Knowledge-based modeling incorporates empirical methods or first principles to improve model generalization. For neural nets, the knowledge can be any extra information about the modeled function beyond the function's inputs/outputs used as training samples~\cite{bandler1994space}. Several studies have shown that the knowledge-based neural nets exhibit better extrapolation capabilities while require fewer training data, compared with vanilla neural nets. In~\cite{stewart2017label}, the neural net is trained by learning a loss function capturing a physical constraint expressed in closed form. This method is also applied in neural surrogate modeling for fluid flows without using any simulator-generated data~\cite{sun2020surrogate}.

This paper aims to develop a surrogate-assisted calibration approach for data center CFD models with high computational cost. We will advance the methodology of designing surrogate to improve its learning efficiency by incorporating first principles and prior knowledge of the modeled data hall. Therefore, we can achieve high calibration performance with much less CFD-generated training data for the surrogate. In addition, to the best of our knowledge, we are the first demonstrating the use of neural surrogate to calibrate industry-grade CFD models for large-scale data halls.

\section{Problem Formulation and Approach Overview}
\label{sec3}

\begin{figure}[t]
	\centering
	\includegraphics[width=1\columnwidth]{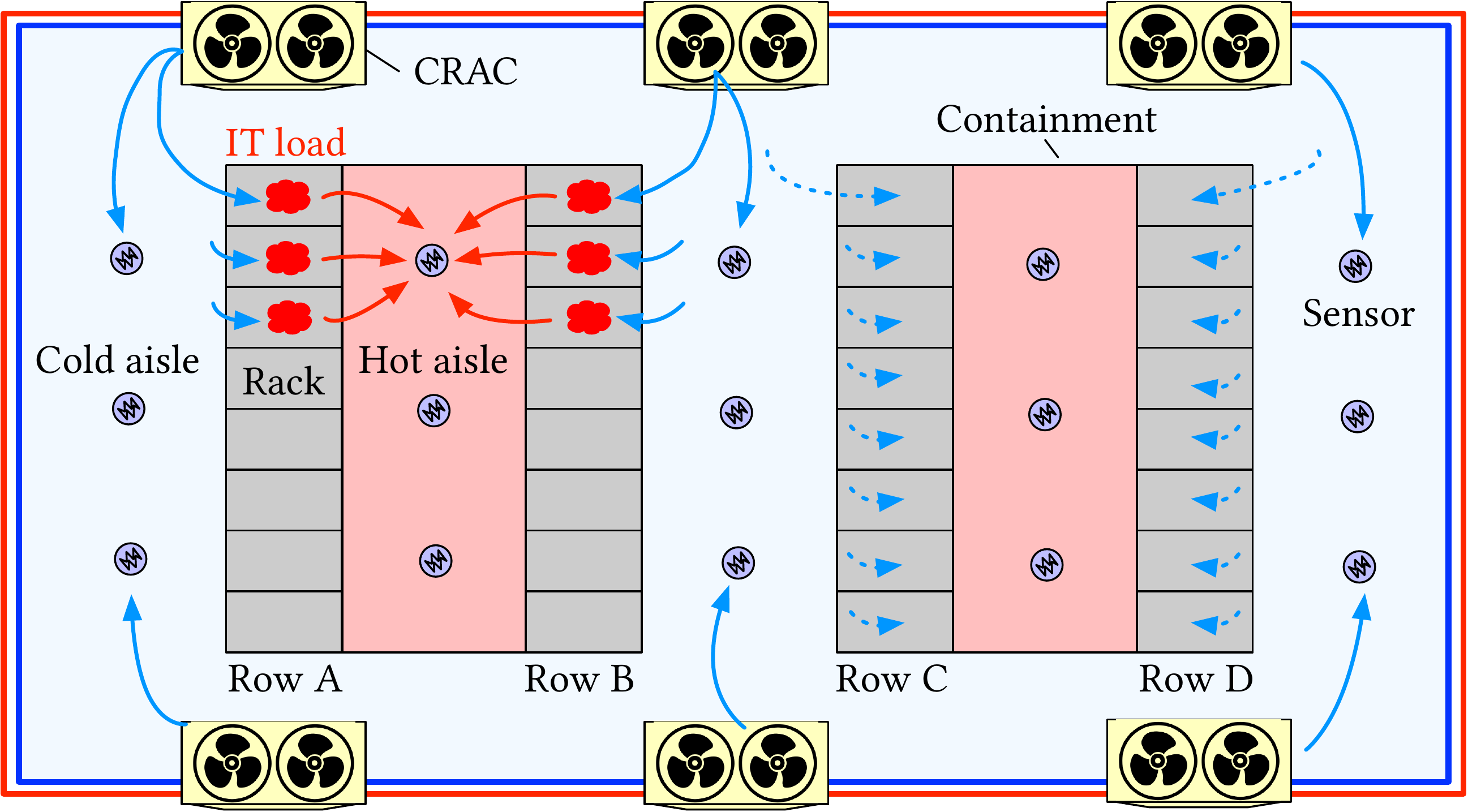}
	\caption{The layout of a typical data hall. Sensors are installed at the cold and hot aisles for cooling evaluation. Sensor measurements are mostly affected by the nearby CRACs.}
	\label{fig2}
\end{figure}

In this section, we introduce the related background. Then, we formulate the problem and present the overview of our approach.

\subsection{Background}
\label{background}
CFD model can estimate the temperature and air velocity distributions in a given space by solving a simplified form of the Navier-Stokes equations~\cite{anderson1995computational}. For air-cooled data centers, CFD has been widely used during the prototyping phase for thermal and air flow analysis to avoid operational risks. To pursue higher efficiency of the cooling systems while not compromise the thermal safety of the computing and network equipment, it is desirable to improve the accuracy of the CFD model toward the paradigm of digital twin in the operational phase of data centers.

\begin{figure}
	\minipage{0.23\textwidth}
	\includegraphics[width=\linewidth]{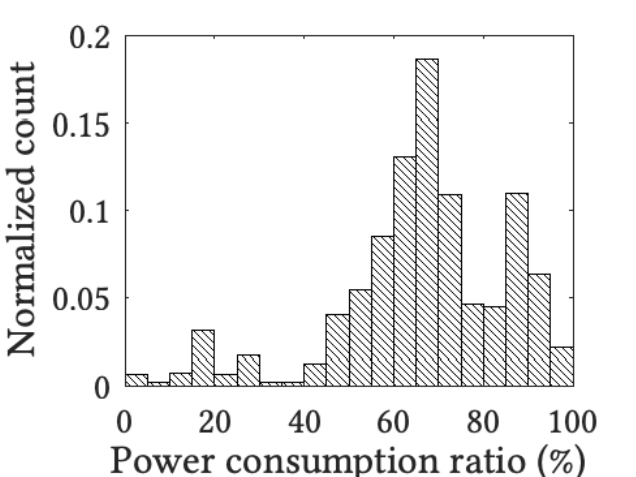}
	\caption{Server power consumption distribution}
	\label{fig3}
	\endminipage\hfill
	\minipage{0.23\textwidth}
	\includegraphics[width=\linewidth]{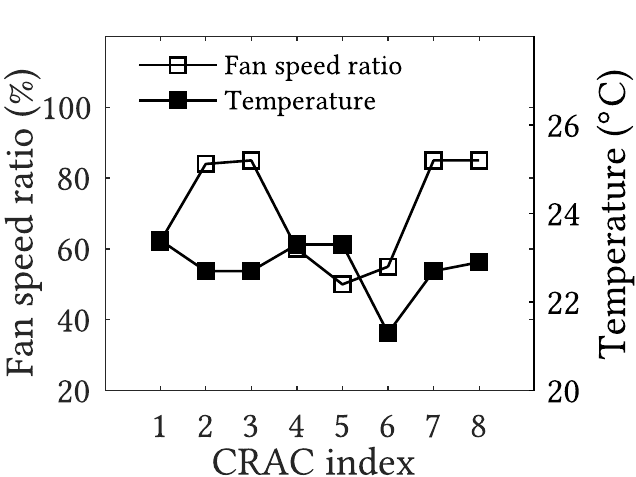}
	\caption{CRAC setpoints and fan speeds}
	\label{fig4}
	\endminipage\hfill
\end{figure}

\begin{table}[t]
	\caption{Summary of Notations.}
	\label{tab1}
	\small
	\begin{tabular}{ll|ll}
		\toprule
		\textbf{Sym.} & \textbf{Definition} & \textbf{Sym.} & \textbf{Definition} \\
		\midrule
		$||\cdot||_2$ & $\ell_2$-norm & $\textbf{e}$ & sensor one-hot vector  \\
		$\otimes$ & Element-wise product & $\textbf{T}_\textup{c}$ & setpoints vector \\
		$l$ & CRAC count & $ \textbf{V} $ & fan speeds vector \\
		$m$& server count & $\textbf{P}$ & powers vector \\
		$n$ & sensor count & $\boldsymbol{\upalpha}$ & flow rates vector \\
		$\alpha_\textup{l}$, $\alpha_\textup{u}$ & $\alpha$ lower and upper bound & $\textbf{T}_\textup{s}$ & measurements vector \\
		$\mathcal{L}_1$ & first loss function & $\widetilde{\textbf{T}}_\textup{s}$ & CFD results vector \\
		$\mathcal{L}_2$ & second loss function & $\widehat{\textbf{T}}_{\textup{s}}$ & surrogate results vector \\
		$\textbf{W}^{\textup{cs}}$ & CRAC to sensor matrix & $\textbf{W}^{\textup{ss}}$ & server to sensor matrix \\
		\bottomrule
	\end{tabular}
\end{table}

Fig.~\ref{fig2} illustrates the layout of a typical data hall, where {\em racks} hosting {\em servers} are assigned into multiple {\em rows} that separate {\em aisles}. These aisles alternate between cold and hot aisles. The {\em computer room air conditioning} units (CRACs) supply cold air to the servers through the cold aisles and draw hot air from the hot aisles. To avoid air recirculation, containments are often implemented for the hot aisles. To evaluate the thermal condition in a data hall, the inlet and outlet temperatures of servers are often used as the key thermal variables. Therefore, temperature sensors are deployed in the cold and hot aisles to monitor such thermal variables. The inlet temperatures are often required to be in the range of 15\textdegree{}C to 27\textdegree{}C \cite{6322262}. The outlet temperatures characterize the heat generated by the servers. Although the CFD model can predict the temperature at any location, we focus on the locations that are deployed with temperature sensors and thus have ground-truth temperature measurements for accuracy evaluation.

The servers in general have different characteristics in passing the cooling air through them. The characteristic highly depends on the server form factor and the control logics of the server's internal fans. Owing to the distinct characteristics, the servers will have different passing-through air flow rates in cubic feet per minute watt (cfm/W), where the cubic feet is for air volume, the minute is for time, and the watt is for the server power. The collection of the server air flow rates is part of the system configuration that greatly affects the thermodynamics of the data hall. Therefore, to achieve high CFD accuracy, the server air flow rates should be configured in the CFD model. Unfortunately, they are often unknown and hard to obtain. The manual {\em in situ} measurement using an air volume flow rate meter for each server is extremely labor intensive, especially for a large-scale data hall that hosts many types of servers. As a result, the server air flow rates are often empirically estimated by human expert. For a CFD model with many (e.g., thousands) servers, the rough settings of the server air flow rates could significantly downgrade the temperature prediction capability of the CFD model. The low accuracy will impede the use of CFD model for the desired fine-grained operational adjustment to pursue energy efficiency without causing thermal risk. 

In this paper, we focus on devising an automatic approach to calibrate the server air flow rates configuration for data center CFD models on a steady system state. The approach can be also extended to include other parameters (e.g., by-pass air flow rates and recirculated air flow rates) into calibration. The system state consists of the following measurements: the setpoints and fan speeds of CRAC units, server powers and the temperatures measured in the hot and cold aisles. With the calibrated server air flow rates, the CFD model will yield more accurate temperature distribution prediction.

\begin{figure}
	\centering
	\includegraphics[width=1\columnwidth]{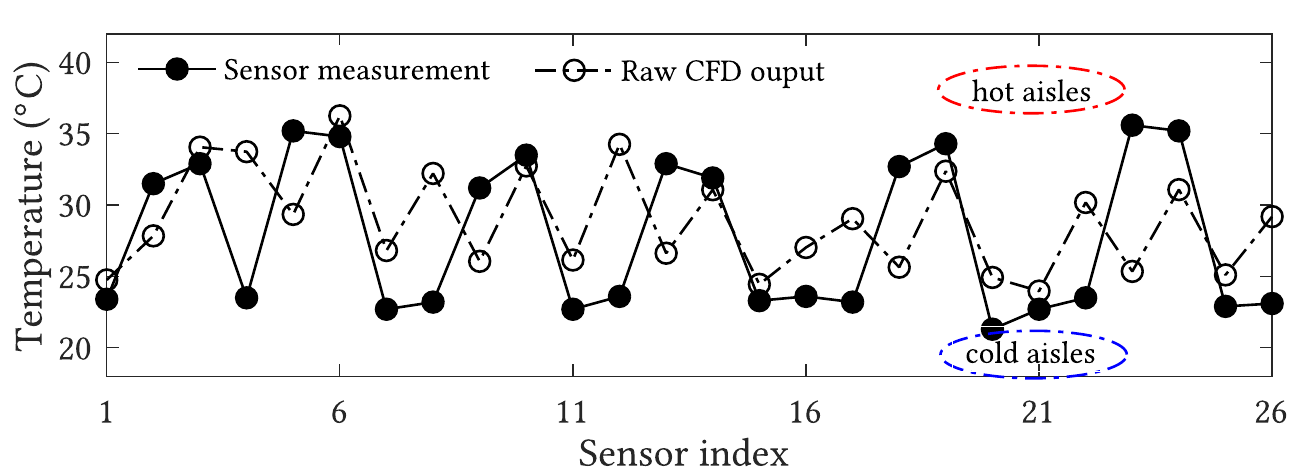}
	\caption{Sensor measurements and raw CFD temperature outputs at corresponding sensor locations.}
	\label{fig5}
\end{figure}

\subsection{Problem Formulation}
\label{3-2}

To formulate the calibration problem, we first define the relevant parameters in a data hall. Unless particularly specified, the notations used in this paper are summarized in Table~\ref{tab1}. We consider a data hall hosting $l$ CRACs, $m$ servers, and $n$ temperature sensors deployed in the cold and hot aisles, respectively. 

\begin{definition}[Input]
	The input data for solving a CFD model is a vector consisting of all modeling parameters. Formally, the input $\textbf{x} = (\textbf{T}_\textup{c}, \textbf{V}, \textbf{P},  \boldsymbol{\upalpha})$, where $\textbf{T}_\textup{c}=(T_{\textup{c}1}, T_{\textup{c}2}, \ldots,T_{\textup{c}l})$, $\textbf{V}=(V_1, V_2, \ldots, V_l)$, $\textbf{P}=(P_1, P_2, \ldots, P_m)$, and $\boldsymbol{\upalpha} = (\alpha_1, \alpha_2, \ldots, \alpha_m)$ are the vectors of CRAC setpoints, CRAC fan speeds, server powers, and server air flow rates, respectively.
	\label{def:1}
\end{definition}

\begin{definition}[Output]
	The output of CFD is a steady-state temperature and air velocity distribution map. For CFD model calibration, we focus on a set of results within the map at the locations installed with temperature sensors, which is denoted by $\widetilde{\textbf{T}}_\textup{s}=(\tilde{T}_{\textup{s}1}, \tilde{T}_{\textup{s}2}, \ldots, \tilde{T}_{\textup{s}n})$.
\end{definition}

\begin{definition}[Measurement]
	The measurement is a vector of real temperature values recorded by the physical sensors, which is denoted by $\textbf{T}_\textup{s}=(T_{\textup{s}1}, T_{\textup{s}2}, \ldots, T_{\textup{s}n})$.
\end{definition}

Let $||\cdot||_2$ denote the $\ell_2$-norm of a vector. With the above definitions, the CFD model calibration aims to find the server air flow rate configuration that minimizes the $\ell_2$-norm of the error vector between the model output and the measurement:
\begin{equation}
	\label{eq1}
	\boldsymbol{\upalpha}^*\! \triangleq\! \operatorname*{arg\,min}_{\boldsymbol{\upalpha}} ||\widetilde{\textbf{T}}_{\textup{s}}(\textbf{x}) - \textbf{T}_{\textup{s}}||^2_2, \! \quad \! \mathrm{s.t.} \; \alpha_\textup{l} \leq \alpha_i \leq \alpha_\textup{u}, i=1,\dots,m,
\end{equation}
where $\boldsymbol{\upalpha}^*$ is the vector of calibrated air flow rates. Each element in $\boldsymbol{\upalpha} ^*$ should be within an empirically estimated range $[\alpha_\textup{l}, \alpha_\textup{u}]$. The servers of the same type in general have the same air flow rate. 

We now use an example in a real production data hall to illustrate the discrepancy of an uncalibrated CFD model and the actual sensor measurements. We first show a summary of the working conditions of the data hall. Fig.~\ref{fig3} shows a sample distribution of the servers' power consumption ratios at a time instant. We can see that most servers are working at approximately $60\%$ of its maximum power. Fig.~\ref{fig4} is the CRAC setpoints and the corresponding fan speed ratios. Fig.~\ref{fig5} shows the temperature values measured by a number of sensors and uncalibrated CFD predictions on the locations of these sensors. For the sensor measurements, the cold aisle temperatures range from 20\textdegree{}C to 24\textdegree{}C, which are related to the CRAC setpoints and fan speed ratios. The hot aisle temperatures range from 30\textdegree{}C to 36\textdegree{}C, which are affected by the generated heat from the servers. The air flow rate of each server is empirically determined for the raw CFD model. With these initial configurations, the CFD model has temperature prediction errors from 2\textdegree{}C to 10\textdegree{}C. Such large errors disqualify the raw CFD model as a data center digital twin.

\begin{figure*}[t]
	\centering
	\includegraphics[width=2\columnwidth]{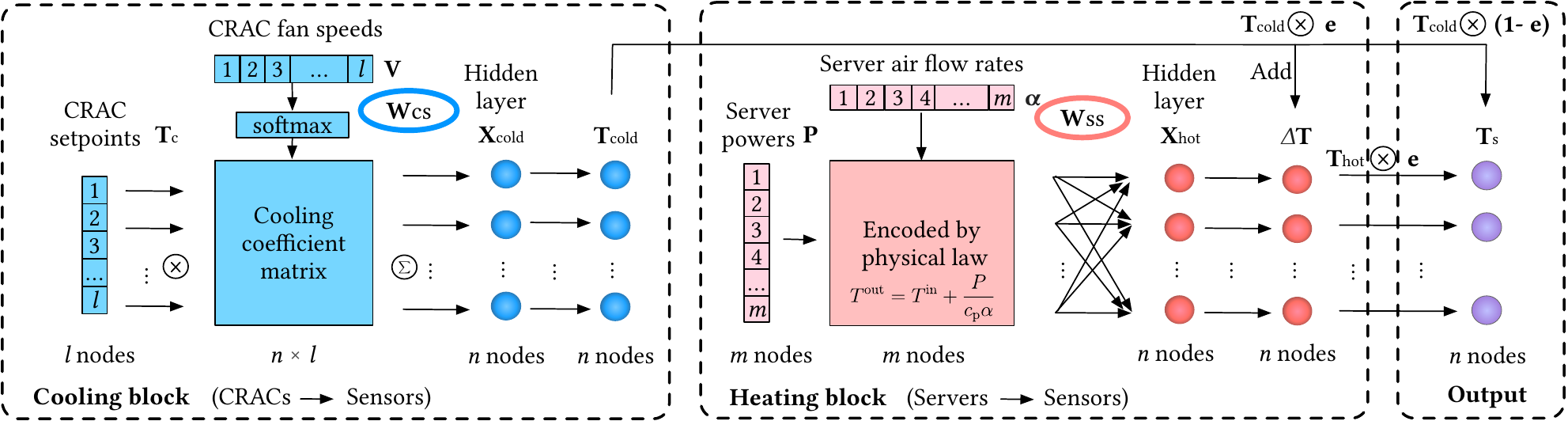}
	\caption{The architecture of the knowledge-based neural surrogate for temperature prediction. The structure consists of a {\em cooling block} and a {\em heating block}. The weight between any two facilities is initialized using their normalized reciprocal spatial distance. Two linear hidden layers are used to predict the cold aisle temperatures and temperature rises induced by servers.}
	\label{fig6}
\end{figure*}

\subsection{Approach Overview}
\label{sec3-3}

Due to the high computational cost of CFD model solving, directly solving the optimization problem in Eq.~(\ref{eq1}) using search algorithms will incur unacceptable computation overhead. To address this issue, we design a surrogate model of the CFD model. Let $\widehat{\textbf{T}}_{\textup{s}} \in \mathbb{R}^{1 \times n}$ denote the temperature output vector of the surrogate model. Then, the problem in Eq.~(\ref{eq1}) is converted to a surrogate-assisted optimization that can be solved by iterating four consecutive steps. First, the surrogate model is trained to be locally aligned with the CFD model by minimizing the discrepancy between the surrogate's and the CFD's outputs:

\begin{equation}
	\begin{aligned}
		\textbf{W}^* \triangleq \operatorname*{arg\,min}_{\textbf{W}}
		||\widetilde{\textbf{T}}_{\textup{s}}(\textbf{x})-\widehat{\textbf{T}}_{\textup{s}}(\textbf{W, x})||^2_2,
	\end{aligned}
\end{equation}

where $\textbf{W}$ is a set of trainable weights of the surrogate and $\textbf{W}^*$ is the result of the surrogate training. Second, with $\textbf{W}^*$, the surrogate is re-optimized through re-training such that the discrepancy between the surrogate's output and the measurement is minimized:

\begin{equation}
	\label{eq3}
	\begin{aligned}
		\boldsymbol{\upalpha}^* \triangleq \operatorname*{arg\,min}_{\boldsymbol{\upalpha}} ||\widehat{\textbf{T}}_{\textup{s}}(\textbf{W}^*, \textbf{x}) -  \textbf{T}_{\textup{s}}||^2_2. \\
	\end{aligned}
\end{equation}

Third, the $\boldsymbol{\upalpha}^*$ is configured into the CFD model. Finally, the CFD is validated based on sensor measurements. If the surrogate approaches to the CFD model, the $\boldsymbol{\upalpha} ^*$ after the convergence of the four-step iterations will approach to the one given by Eq.~(\ref{eq1}). 

As discussed in \cref{sec1}, to address the challenges of the surrogate's complexity versus the needed volume of CFD-generated training data, we build a knowledge-based neural surrogate that can capture the physical layout and thermal relations among a number of key variables of the considered data hall. Specifically, we model a set of facilites (i.e., CRACs, servers, and sensors) in the considered hall as nodes and their connections as edges into a directed graph. The direction of an edge characterizes the thermal causality between the two end nodes of the edge. For example, an edge points from a CRAC node to a sensor node, because the supply air temperature of the CRAC affects the measured temperature of the sensor.  The normalized reciprocal spatial distance between any two facilities will be used as the weight of the edge connecting the corresponding two nodes in the graph. This modeling approach follows the fact that the temperature measured by a sensor is mostly affected by the facilities in its neighborhood~\cite{lazic2018data}.

\section{CFD Calibration via Surrogate}

\label{sec4}
In this section, we present the design of the knowledge-based neural surrogate and Kalibre's iterative four-step model calibration.

\subsection{Knowledge-based Neural Surrogate}

The neural surrogate aims to approximate the complex thermophysics encompassed in the CFD model.
In particular, its efficient training with a small amount of data generated from the CFD model is desirable, since the data generation requires intensive computation. Fig.~\ref{fig6} shows the proposed neural surrogate architecture. It consists of a {\em cooling block} and a {\em heating block}. The cooling block models the impact of the CRACs on the temperatures at all sensor locations; the heating block models the impact of the servers on the temperatures at the hot aisle sensor locations. Thus, the sum of the two blocks captures the effects from both CRACs and servers.
The input of the model consists of the free variables of the data hall's steady state at a time instant, including CRAC temperature setpoints and fan speeds, and server powers. The server air flow rates are designated as trainable variables of the neural surrogate and initialized with rough estimates. Note that, as the neural surrogate is differentiable, the server air flow rates can be updated efficiently by backpropagation-based neural net training algorithms for the purpose of calibration.
The output of the neural surrogate is a vector of $n$ predicted temperatures at the sensor locations. In what follows, we present the designs of the cooling and heating blocks of the neural surrogate to capture prior knowledge of the thermal relations among the key variables. Lastly, we present the settings of the constants used by the neural surrogates, which are also based on the prior knowledge on the layout of the modeled data hall.

\subsubsection{Cooling block}
This block models the impact of the CRAC temperature setpoints $\textbf{T}_\textup{c}$ and fan speeds $\textbf{V}$ on the temperatures at all sensor locations.
First, we encode the two free variables (i.e., $\textbf{T}_\textup{c}$ and $\textbf{V}$) into a hidden-layer variable for the $k^{\textup{th}}$ sensor as
$X^{\textup{cold}}_k = \sum_{i=1}^{l} T_{\textup{c}i} \cdot c_{ik}$, where
$T_{\textup{c}i} $ is the setpoint of the $i^{\textup{th}}$ CRAC and $c_{ik}$ is a cooling coefficient characterizing the impact of the $i^{\textup{th}}$ CRAC on the $k^{\textup{th}}$ sensor. We design $c_{ik}$ to be positively related to the CRAC fan speed. Specifically, we use softmax activation to compute the {\em cooling coefficient matrix} as $c_{ik} = \frac{e^{z_{ik}}}{\sum_{a=1}^{l}e^{z_{ak}}}$,
where $z_{ik}$ is an intermediate variable defined by $z_{ik} = V_i \cdot W^{\textup{cs}}_{ik}$, $V_i$ is the fan speed of the $i^{\textup{th}}$ CRAC, and $W_{ik}^{cs}$ is a weight characterizing the thermal impact of the $i^{\textup{th}}$ CRAC on the $k^{\textup{th}}$ sensor. The CRAC-to-sensor matrix $\textbf{W}^{\textup{cs}} \in \mathbb{R}^{n \times l}$ consisting of $W_{ik}^{cs}$ for $i = 1, \ldots, l$ and $k = 1, \ldots, n$ is an adjacency matrix. The weights in this matrix can be fixed or trainable. If they are fixed, their settings are important and will be discussed in \cref{subsubsec:connection-weights}. \cref{sec5} will compare the performance of the neural surrogates with $\textbf{W}^{\textup{cs}}$ fixed or trainable.
Lastly, we use a linear layer to project the hidden-layer variable to temperature as $T^{\textup{cold}}_k = a_k X^{\textup{cold}}_k + b_k$,
where $a_k$ and $b_k$ are two trainable weights.

\subsubsection{Heating block}

This block models the impact of the servers on the temperatures at the hot aisle sensor locations. We assume that the energy dissipated from the servers in the forms of electromagnetic radiation and mechanical movements is negligible compared with that dissipated in the form of heat. Thus, from \cite{6322262}, the temperature increase caused by a server consuming $P$ watts at its outlet can be modeled by $\frac{P}{c_{\textup{p}}\alpha}$, where $c_{\textup{p}}$ is a heating constant representing the heat capacity of air and $\alpha$ is the server air flow rate. Based on this first principle, we use server powers and air flow rates to predict the server-induced temperature increase $\Delta T_k$ at the $k^{\textup{th}}$ hot aisle sensor location. Specifically, $\Delta T_k  = c_k X^{\textup{hot}}_k + d_k$, where $c_k$ and $d_k$ are two trainable weights, and $X^{\textup{hot}}_k$ is a hidden-layer variable. The $X^{\textup{hot}}_k$ is defined by
$X^{\textup{hot}}_k=\sum_{j=1}^{m} \frac{P_j}{\alpha_j} \cdot W^{\textup{ss}}_{jk}$,
where $P_j$ is the $j^{\textup{th}}$ server power, $\alpha_j$ is the $j^{\textup{th}}$ server air flow rate, and $W_{ij}^{ss}$ is the weight
characterizing the thermal impact of the $j^{\textup{th}}$ server on the $k^{\textup{th}}$ sensor. We define the server-to-sensor adjacency matrix $\textbf{W}^{\textup{ss}}$ consisting of $W^{\textup{ss}}_{jk}$ for $j = 1, \ldots, m$ and $k = 1, \ldots, n$. Similar to $\textbf{W}^{\textup{cs}}$, $\textbf{W}^{\textup{ss}}$ can be fixed or trainable. For the former case, its setting is discussed in \cref{subsubsec:connection-weights}. Note that the heating block outputs $\Delta T_k$ for all sensor locations (denoted by $\Delta \textbf{T}$); but only the outputs at hot aisle sensor locations will be used when combing the results of the cooling and heating blocks. This design simplifies the vectorized implementation of the neural surrogate using TensorFlow (cf.~\cref{4-3}).

\subsubsection{Joining two blocks and adjacency matrices settings}
\label{subsubsec:connection-weights}
With hot-aisle containment and blanket, heat recirculation is negligible. Thus, the temperatures at cold aisle sensor locations are mainly affected by the CRACs; the temperatures at hot aisle sensor locations are jointly affected by the CRACs and servers. To combine the outputs of the cooling and heating blocks, we define a one-hot vector $\textbf{e} \in \{0,1\}^n$, where its element $e_k=1$ or $0$ represents that the $k^{\textup{th}}$ sensor location is in hot or cold aisle, respectively. Therefore, the final output of the neural surrogate, i.e., the temperatures at all sensor locations, can be expressed by $\widehat{\textbf{T}}_{\textup{s}} = \textbf{T}_{\textup{cold}} \otimes (\boldsymbol{1} - \textbf{e}) + \textbf{T}_{\textup{cold}} \otimes \textbf{e} + \Delta \textbf{T} \otimes (\boldsymbol{1} - \textbf{e})$, where $\otimes$ represents element-wise product, $\textbf{T}_{\textup{cold}} \otimes (\boldsymbol{1} - \textbf{e})$ gives the temperatures at the cold aisle sensor locations, and $\textbf{T}_{\textup{cold}} \otimes \textbf{e} + \Delta \textbf{T} \otimes (\boldsymbol{1} - \textbf{e})$ gives the temperatures at the hot aisle sensor locations.

If $\textbf{W}^{\textup{cs}}$ and $\textbf{W}^{\textup{ss}}$ are fixed, the weights of the neural surrogate are $\textbf{W} = \{a_k, b_k, c_k, d_k | k = 1, \ldots, n\}$; otherwise, $\textbf{W}$ additionally include $\textbf{W}^{\textup{cs}}$ and $\textbf{W}^{\textup{ss}}$. We now discuss the settings of $\textbf{W}^{\textup{cs}}$ and $\textbf{W}^{\textup{ss}}$ if they are not trainable. Each of their elements represents the thermal impact of a facility (CRAC or server) on a sensor location. Since the thermal impact decreases with spatial distance, in this paper, we set it to be a normalized reciprocal of the spatial distance between the facility and the sensor location. When it is lower than a threshold, it is forced to be zero, indicating that the corresponding thermal impact is negligible.
Thus, to set these two two matrices, the layout of the data hall and the sensor locations will be needed, which are available to the data center operator in general.

\subsection{Four-step Iterations for CFD Calibration}
\label{4.2}

Let $\widehat{T}_{\textup{s}k}$, $\widetilde{T}_{\textup{s}k}$, $T_{\textup{s}k}$ denote the surrogate-predicted temperature, CFD-predicted temperature, and the measured temperature at the location of the $k^{\textup{th}}$ sensor, respectively. Algorithm~\ref{Alg1} shows the pseudocode of the four-step iterations. We now explain it in detail.

{\bf \ding{172} Neural surrogate training} (Line~\ref{line:4}-\ref{line:6}): The training data is generated by solving the CFD model with collected system input (including CRAC temperature setpoints and fan speeds, server powers) and the initial  $\boldsymbol{\upalpha}$ or calibrated $\boldsymbol{\upalpha}$ by step \ding{174} of the previous iterations to yield the predicted temperatures at sensor locations. The detailed training data generation is described in \cref{sec5-3}. Note that each element of $\boldsymbol{\upalpha}$ should be within  $[\alpha_l, \alpha_u]$. The system input, the $\boldsymbol{\upalpha}$, and the predicted temperatures form a new training data sample that is added to the training dataset accumulated from the first iteration. With the training dataset, the neural surrogate is updated to minimize the errors between its predicted temperatures and the CFD-predicted temperatures of the training samples. Thus, the weights of the neural surrogate are updated using the gradient of the least squares loss function of $\mathcal{L}_1 = \frac{1}{n} \sum_{k=1}^{n} (\widehat{T}_{\textup{s}k}(\textbf{W},\textbf{x}) - \widetilde{T}_{\textup{s}k}(\textbf{x}))^2$.
As a result, the surrogate is trained to align with the CFD model. At the end of this step, $\textbf{W}$ is frozen. 

\begin{algorithm}[t]
	\caption{Kalibre's CFD model calibration procedure.}
	\label{Alg1}
	\begin{algorithmic}[1]
		\REQUIRE {Measurements collected from a data hall at a time instant, including CRAC setpoints $\textbf{T}_{\textup{c}}$, CRAC fan speed ratios $\textbf{V}$, server powers $\textbf{P}$, and sensor measurements $\textbf{T}_{\textup{s}}$. Initial server air flow rates $\boldsymbol{\upalpha}$. Cooling and heating coefficient matrix $\textbf{W}^{\textup{cs}}$ and $\textbf{W}^{\textup{ss}}$.}
		\ENSURE {Calibrated $\boldsymbol{\upalpha}$.}
		\STATE {Initialize each $\alpha$ within $[\alpha_\textup{l}, \alpha_\textup{u}]$  and CFD error $\epsilon$;}
		\STATE {Assign initial configurations to the surrogate graph $\mathcal{G}$;}
		\FOR {$i=1$ : Max iteration}
		\STATE {Solve CFD model to obtain $\widetilde{\textbf{T}}_{\textup{s}}$;}
		\label{line:4}
		\STATE {Aggregate CFD solving results as training data;}
		\STATE {Train surrogate by performing gradient descent on $\mathcal{L}_1$;}
		\label{line:6}
		\STATE {Search $\boldsymbol{\upalpha}$ by performing differential evolution;}
		\label{line:7}
		\STATE {Search $\boldsymbol{\upalpha}$ by performing gradient descent on $\mathcal{L}_2$;}
		\label{line:8}
		\STATE {Configure $\boldsymbol{\upalpha}$ to the CFD model;}
		\label{line:9}
		\IF {$\frac{1}{n}\sum_{i}^{n}| \widetilde{T}_{\textup{s}i} - T_{\textup{s}i}| < \epsilon$}
		\label{line:10}
		\STATE {$\epsilon \leftarrow \frac{1}{n}\sum_{i}^{n}| \widetilde{T}_{\textup{s}i} - T_{\textup{s}i}|$; $\quad$ $\boldsymbol{\upalpha}^* \leftarrow \boldsymbol{\upalpha}$;}
		\ENDIF
		\label{line:13}
		\ENDFOR
		\RETURN {Calibrated server air flow rate configurations $\boldsymbol{\upalpha}^*$;}
	\end{algorithmic}
\end{algorithm}

{\bf \ding{173} Surrogate-assisted calibration} (Line~\ref{line:7}-\ref{line:8}): The surrogate is re-optimized to minimize the errors between its predicted temperatures and the measured temperatures by updating $\boldsymbol{\upalpha}$. In this step, $\boldsymbol{\upalpha}$ is set trainable. An empirical regularization term is added to penalize the loss function if the temperature difference between the hot and cold aisle, i.e., $\Delta T$, is out of the empirical range [$\Delta T_{\textup{l}}, \Delta T_{\textup{u}}$]. The penalty term is expressed using the rectified linear units (ReLU) as $h(T) = \sum_{j=1}^{m} (\text{ReLU}(\Delta T_{\textup{l}} - \Delta T) + \text{ReLU}(\Delta T - \Delta T_{\textup{u}})) \times P_j$, where $P_j$ is the $j^{\text{th}}$ server power. The term means that, if the server power is higher, the penalty should be more significant. Thus, the second loss function with regularization is $\mathcal{L}_2 = \frac{1}{n} \sum_{k=1}^{n} (\widehat{T}_{\textup{s}k}(\boldsymbol{\upalpha}) - T_{\textup{s}k})^2 + \frac{\lambda}{n}  \sum_{k=1}^{n}h(T)$, where $\lambda$ is a regularization coefficient. In our experiments, we set $\Delta T_{\textup{l}}$ and $\Delta T_{\textup{u}}$ to be 5\textdegree{}C and 15\textdegree{}C, based on the data center operator's experience. To accelerate the re-optimization, we implement a hybrid approach of combining differential evolution algorithm with gradient backpropagation to minimize the loss function $\mathcal{L}_2$. This hybrid approach has been shown effective in accelerating neural net training~\cite{wang2015back}. We will also evaluate its effectiveness for our specific problem in \cref{sec5-3}.

{\bf \ding{174} CFD configuration} (Line~\ref{line:9}): The updated $\boldsymbol{\upalpha}$ is configured back to the CFD model. The refined CFD model is then used for step \ding{172} of the next iteration.

{\bf \ding{175} CFD validation} (Line~\ref{line:10}-\ref{line:13}): The CFD model's accuracy is validated against the ground-truth sensor measurements. Only better $\boldsymbol{\upalpha}$ is recorded for final output candidate.

Through iterative optimization of the two loss functions $\mathcal{L}_1$ and $\mathcal{L}_2$, the $\boldsymbol{\upalpha}$ will be calibrated to improve the CFD model's accuracy.

\subsection{Implementation of Kalibre}
\label{4-3}

We implement Kalibre with Python 3.5 and Google TensorFlow 1.15.0, where the latter is a library widely used for building machine learning applications. When we use TensorFlow to build the neural surrogate's computational graph, the server air flow rates $\boldsymbol{\upalpha}$ are set as a vector of trainable variables instead of a TensorFlow placeholder. This allows us to control their updating by choosing to freeze the gradients or not. We choose Adam \cite{kingma2014adam} as the optimizer, which is a method for efficient stochastic optimization that only requires first-order gradients and little memory space. The CFD model solving is performed by 6SigmaDCX \cite{6Sigma}, a commercial CFD software package. The 6SigmaDCX can load $\boldsymbol{\upalpha}$ from a configuration file. During the four-step iterations, our Python program writes the candidate $\boldsymbol{\upalpha}$ to the file, invokes a 6SigmaDCX session to solve the CFD model, and collects results by parsing 6SigmaDCX's output.

\section{Performance Evaluation}
\label{sec5}
In this section, we apply Kalibre to calibrate the CFD models built for two production data halls and present the evaluation results against other baseline approaches.

\begin{table}
	\caption{CFD model solving time.}
	\label{tab2}
	\begin{tabular}{ccccccc}
		\toprule
		CPU cores & 1  & 2  & 4 & 8 & 16 & 32\\
		\midrule
		Solving time (h) & 5.95
		& 3.72 & 2.54 & 0.99 & 0.6 & 0.44 \\
		\bottomrule
	\end{tabular}
\end{table}

\subsection{Experiment Methodology and Settings}

\subsubsection{Data halls and CFD models}
\label{subsubsec:data-halls}

Our targets are two production data halls (referred to as Hall A and Hall B) in operation for e-commerce applications.
Both of them are sized hundreds of square meters that host thousands of servers, respectively (the details of the two data halls are omitted here due to confidentiality requirement). Their CFD models were built and meshed with 10 million grid cells by a domain expert using 6SigmaDCX. The accuracy of these two CFD models will be evaluated in~\cref{sec5-3}. Here, we present their compute overheads. Table~\ref{tab2} shows the compute times for solving one of the CFD models when the number of used CPU cores varies. Note that the 6SigmaDCX software package made available to us supports parallel computing with up to 32 CPU cores on the same computer. The evaluation shows that a single CFD model solving takes up to several hours and the solving time decreases with the number of used CPU cores. However, the model solving speed (i.e., the reciprocal of the solving time) is sub-linear to the number of used CPU cores. This suggests that the CFD computation is not completely divisible and the communications among the paralleled units matter. Thus, even if the 32-core limit is lifted, the attempt to use more CPU cores across multiple computers may face performance bottlenecks due to the cross-computer communication overheads. With 32 CPU cores, the CFD model solving time is about half an hour. This solving time still renders the heuristic search-based model calibration approaches impractical, since they generally need a large number of iterations (e.g., hundreds as shown shortly). Note that GPU acceleration has been introduced to another commercial CFD software package \cite{fluent-gpu}. However, it brings 3.7x acceleration only \cite{fluent-gpu}, which does not change the impracticality of the heuristic search-based model calibration approaches.

\begin{table}[t]
	\caption{Hyperparameter settings of Kalibre.}
	\label{tab3}
	\small
	\begin{tabular}{ll|ll}
		\toprule
		\textbf{Hyperparameter} & \textbf{Setting} & \textbf{Hyperparameter} & \textbf{Setting} \\
		\midrule
		$[\alpha_\textup{l}, \alpha_\textup{u}]$ (cfm/W) & [0.01, 3] & $[\Delta T_{\textup{l}}, \Delta T_{\textup{u}}]$ (\textdegree{}C)& [5, 15] \\
		Augment batch size & 16 & Training epoch & 150 \\
		Initial learning rate & 0.1 & Regularization term & 1 \\
		Decay coefficient & 0.8 & Max search iteration & 100 \\
		Population size & 10 & Crossover rate & 0.6 \\
		\bottomrule
	\end{tabular}
\end{table}

\label{5.2.1}
\begin{figure}[t]
	\centering
	{\includegraphics[width=0.95\columnwidth]{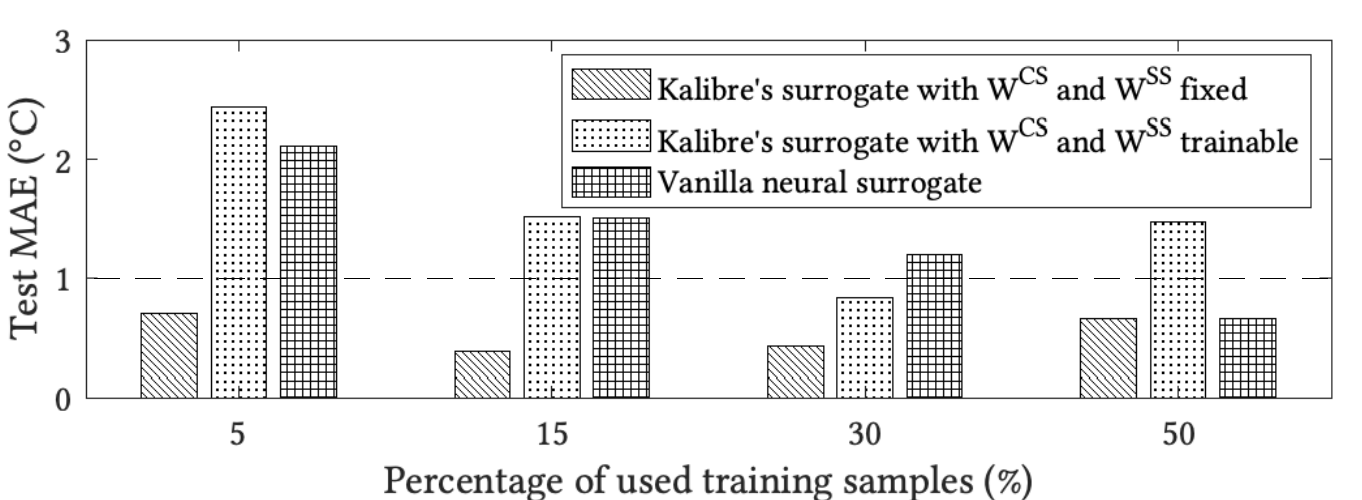}}
	\caption{MAEs of three neural surrogates trained with 5\%, 15\%, 30\%, and 50\% of samples in the training dataset.}
	\label{fig7}
\end{figure}

\subsubsection{Error metric and settings}
We use \emph{mean absolute error} (MAE) to measure the errors of the CFD models in temperature prediction. Specifically, $MAE=\frac{1}{N} \sum_{i=1}^{N}|y_i - \hat{y}_i|$, where $N$ is the number of deployed sensors, $y_i $ and $\hat{y}_i$ are the $i^{\textup{th}}$ sensor measurement and the prediction made by the CFD, respectively. Table~\ref{tab3} shows the hyperparameter settings of Kalibre. These settings include the empirical bounds and the hyperparameters of the surrogate training and differential evolution search. They are selected based on advice from the domain expert or extensive experimental tests. 

\subsection{Evaluation Results}

\subsubsection{Performance of neural surrogates}
As the neural surrogate's efficiency in learning from small data is a key merit, we conduct experiments to investigate the impact of training data volume on the accuracy of the neural surrogate. We consider three designs of the neural surrogate: Kalibre's neural surrogate with $\textbf{W}^{\textup{cs}}$ and $\textbf{W}^{\textup{ss}}$ fixed and trainable, respectively, and a vanilla neural surrogate. The vanilla neural surrogate has three fully-connected layers consisting of 518, 128, and 32 neurons. Before the experiments, we solve Hall A's CFD model to generate 213 training data samples. Each model solving is based on an $\boldsymbol{\upalpha}$ with each of its element sampled randomly and uniformly from $[\alpha_\textup{l}, \alpha_\textup{u}]$.
Then, we divide the generated data samples into training and test datasets following a 8:2 ratio.
Fig.~\ref{fig7} shows the MAEs measured on the test dataset when the three neural surrogates are trained using 5\%, 15\%, 30\%, and 50\% samples of the training dataset.
First, we compare Kalibre's neural surrogates with $\textbf{W}^{\textup{cs}}$ and $\textbf{W}^{\textup{ss}}$ fixed and trainable. We can see that for all amounts of used training data, the neural surrogate with $\textbf{W}^{\textup{cs}}$ and $\textbf{W}^{\textup{ss}}$ fixed and initialized with spatial distance reciprocals outperforms the others. This is because when the two adjacency matrices are given, $a, b, c$ and $d$ are the only parameter sets that we need to learn, i.e. $\textbf{W}$=\{ $a, b, c, d$ \} $\in \mathbb{R}^{4n}$. 
In contrast, the neural surrogate with $\textbf{W}^{\textup{cs}}$ and $\textbf{W}^{\textup{ss}}$ trainable has $(l \times n + m \times n)$ more weights to be learned, which require more training samples to avoid overfitting. Thus, in the rest of this paper, we fix $\textbf{W}^{\textup{cs}}$ and $\textbf{W}^{\textup{ss}}$ at their initial settings based on spatial distance reciprocals. Second, we examine the results of the vanilla neural surrogate. From Fig.~\ref{fig7}, when 5\% training samples are used, the vanilla neural surrogate produces 2.11\textdegree{}C MAE, compared with Kalibre neural surrogate's 0.69\textdegree{}C MAE.
Although the vanilla neural surrogate's MAE decreases with the amount of used training data and eventually achieves comparable MAE as Kalibre's neural surrogate when 50\% training samples are used, the results clearly suggest that the vanilla neural surrogate has lower learning efficiency on small data. This is consistent with our understanding since the vanilla neural surrogate has thousands trainable weights and thus needs more training data to avoid overfitting.

\begin{figure}[t]
	\centering
	{\includegraphics[width=1\columnwidth]{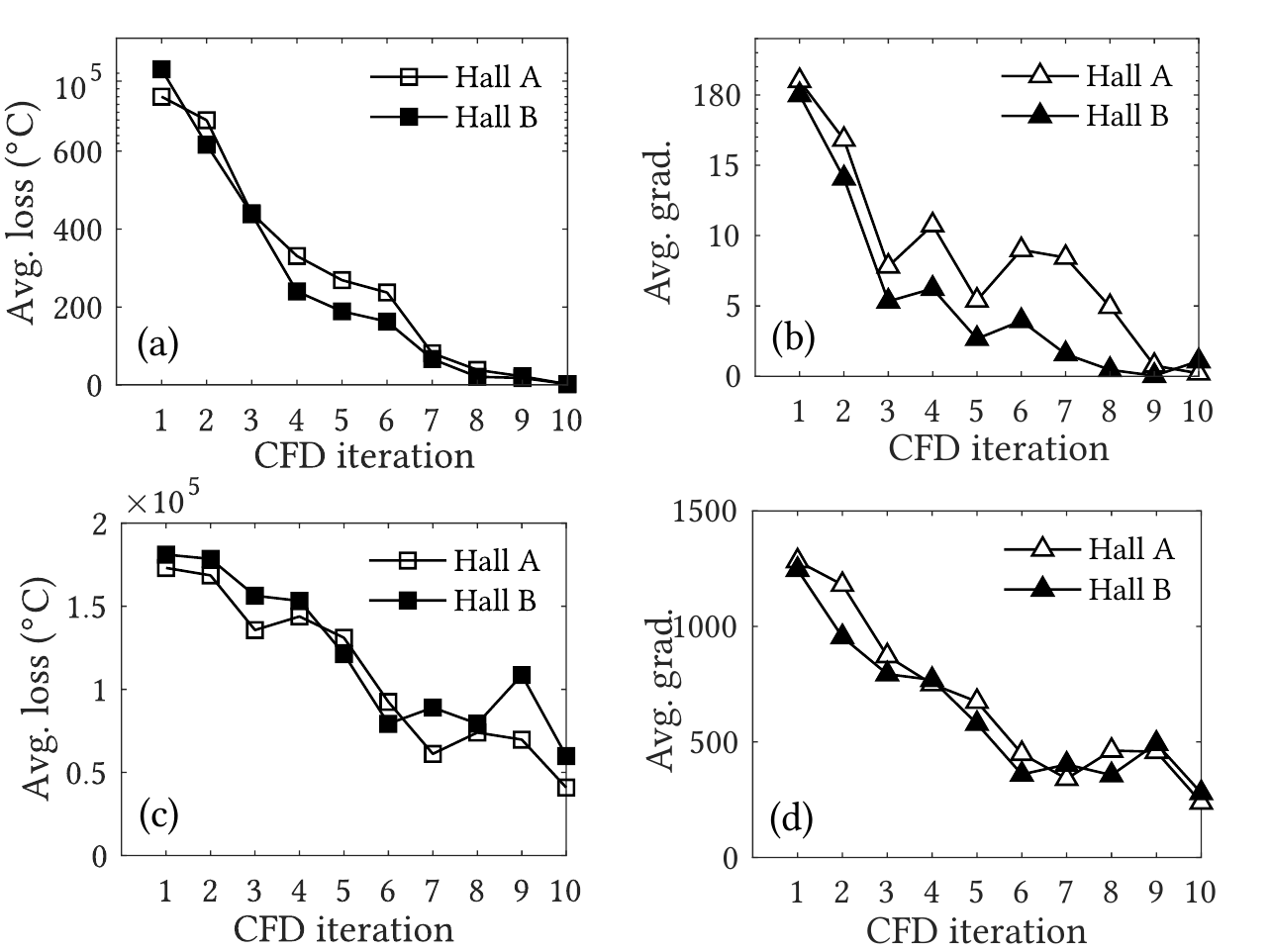}}
	\caption{Average loss and gradient over Kalibre's iterations. (a)-(b): with differential evolution; note that the $y$-axis scale is not uniform. (c)-(d): without differential evolution.}
	\label{fig8}
\end{figure}

\subsubsection{Convergence and effectiveness of Kalibre}
\label{sec5-3}

\begin{figure}[t]
	\centering
	{\includegraphics[width=1\columnwidth]{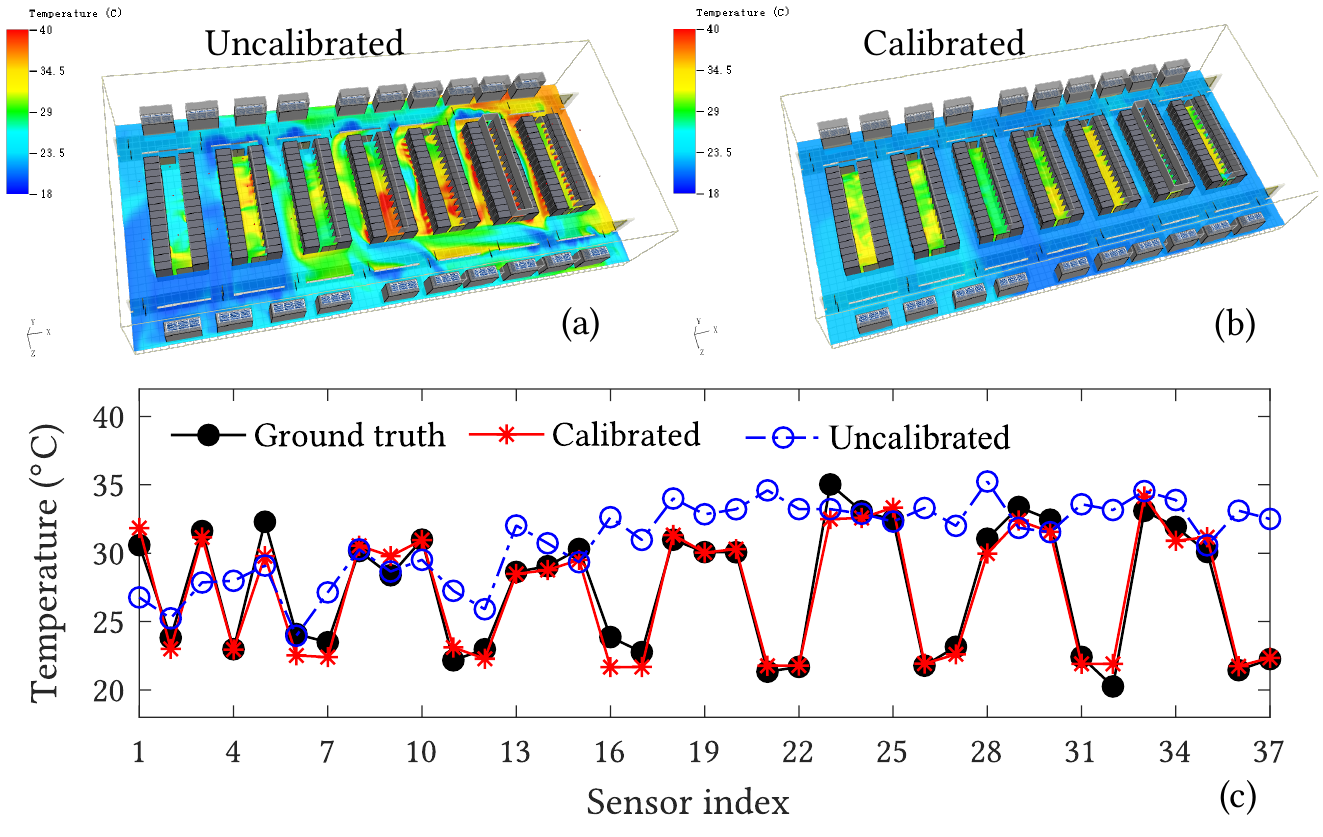}}
	\caption{Hall A temperature distribution. (a) Thermal plane produced by the original CFD model; (b) Thermal plane produced by the calibrated CFD model; (c) CFD-predicted and ground-truth temperatures at the sensor locations.}
	\label{fig9}
\end{figure}

In this set of experiments, we evaluate Kalibre's convergence speed and the effectiveness of its calibration.
To initiate Kalibre's four-step iterations, we solve the CFD model to generate three training data samples by setting each element of $\boldsymbol{\upalpha}$ to the upper and lower bounds, as well as a mid point between the two bounds.
To mitigate the initial overfitting, we augment the three-sample dataset by adding Gaussian noises as in~\cite{doi:10.1162/neco.1995.7.1.108}. The augmented batch size for each sample is $16$. Thus, we have a total of $48$ samples for the initial training process. In each four-step iteration, a new training data sample will be generated by solving the CFD model configured with the $\boldsymbol{\upalpha}^*$ found by the neural surrogate. This new training data sample is aggregated to the training dataset.
In this set of experiment, Kalibre terminates after ten iterations.
As presented in \cref{4.2}, Kalibre adopts a hybrid approach combining the gradient backpropagation widely used for neural net training and the differential evolution to find $\boldsymbol{\upalpha}^*$. In our experiments, the gradient backpropagation is implemented by the Adam optimizer.
Fig.~\ref{fig8}(a) and (b) show the average loss and gradient over the four-step iterations. In the first iteration, the average loss and gradient are very large, reaching around $10^5$ and $180$, respectively.
A closer examination shows that the regularization penalty of the loss function $\mathcal{L}_2$ is large in the very early iterations.
However, in the subsequent iterations, the average loss sharply decreases and converges to zero in the tenth iteration. The average gradient also approaches zero. For comparison, we adopt a baseline of using the Adam optimizer only to find $\boldsymbol{\upalpha}^*$.
Figs.~\ref{fig9}(c) and (d) show the results of this baseline.
We can see that the convergence is slow and the average loss remains large (about $0.5 \times 10^5$) after ten iterations. The results show that the differential evolution effectively accelerates the convergence of Kalibre.

\begin{figure}[t]
	\centering
	{\includegraphics[width=1\columnwidth]{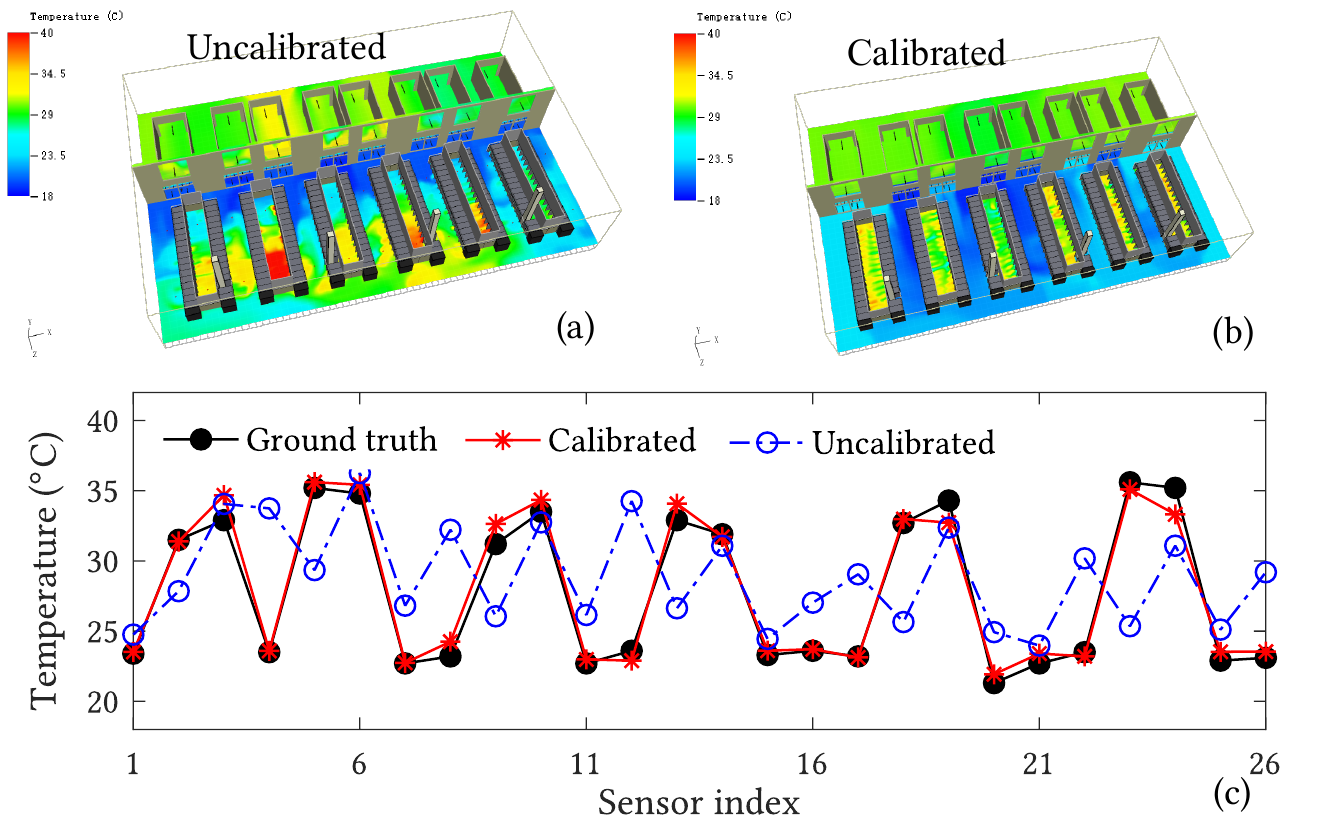}}
	\caption{Hall B temperature distribution. (a) Thermal plane produced by original CFD model; (b) Thermal plane produced by calibrated CFD model; (c) CFD-predicted and ground-truth temperatures at the sensor locations.}
	\label{fig10}
\end{figure}

Then, we show the effectiveness of the model calibration. Figs.~\ref{fig9}(a) and \ref{fig10}(a) show the two halls' thermal planes computed based on the original CFD models presented in \cref{subsubsec:data-halls}. We can see that the temperature distribution is uneven in both the cold and hot aisles. Figs.~\ref{fig9}(c) and \ref{fig10}(c) shows the temperatures predicted by the two halls' original CFD models at the sensor locations and the ground-truth values measured by the sensors. The original CFD models' prediction errors are from 3\textdegree{}C to 6\textdegree{}C. Such large errors are due to the inaccurate estimation of the server air flow rates.
Figs.~\ref{fig9}(b) and \ref{fig10}~(b) show the thermal planes computed based on the CFD models after ten calibration iterations of Kalibre.
We can see that the temperature distributions become more uniform, compared with the results shown in Figs.~\ref{fig9}(a) and Figs.~\ref{fig9}(b).
From Figs.~\ref{fig9}(c) and \ref{fig10}(c), the temperatures predicted by the calibrated CFD models well match the ground-truth values, with MAEs of 0.81\textdegree{}C and 0.75\textdegree{}C for the two halls, respectively. The above results show the effectiveness of Kalibre for large-scale data halls.

\subsubsection{Comparison with baseline approaches}
We compare Kalibre with three baseline approaches discussed in \cref{sec1}:

\vspace{0.2em}
\noindent {\bf \underline{Manual} calibration} involves extensive tuning of the server air flow rates by a CFD expert with years of experience. Specifically, if the CFD-predicted temperature at a sensor location is higher than the ground-truth temperature, the expert empirically increases the flow rates of nearby servers and vice versa.

\vspace{0.2em}
\noindent {\bf \underline{Heuristic} parameter search} uses the covariance matrix adaptation evolution strategy (CMA-ES) to search good $\boldsymbol{\upalpha}$. It solves the CFD model every search iteration. CMA-ES is a gradient-free numerical optimization method. It applies the (1+1) strategy described in \cite{rechenberg1973evolutionsstrategie} to generate one candidate solution per iteration. If the MAE of the new offspring is smaller, it becomes the parent. The mutation rate is set to $\sigma=5$ and updated for each iteration by following the 1/5 successful evolution rule described in \cite{rechenberg1973evolutionsstrategie}.

\vspace{0.2em}
\noindent {\bf \underline{Vanilla} neural surrogate} uses the neural net presented in \cref{5.2.1} that consists of three fully-connected layers. It also follows Kalibre's four-step iterations to perform the model calibration.

Table~\ref{tab4} shows the MAEs achieved by different approaches with ten calibration iterations, as well as the lowest MAEs achieved and the needed iterations.
With ten calibration iterations, Kalibre achieves lower MAEs compared with the baseline approaches for both halls. After about 15 calibration iterations, Kalibre's MAEs converge to 0.76\textdegree{}C and 0.59\textdegree{}C for the two halls. As shown in \cref{5.2.1}, the vanilla neural surrogate requires more training data samples to well represent the CFD model. Thus, with ten calibration iterations, its calibrated CFDs still yield MAEs higher than manual calibration. The heuristic parameter search cannot find good configurations with the same CFD iteration times. Its MAEs saturate at high levels of 3.49\textdegree{}C and 2.61\textdegree{}C even after 120 CFD model solving processes for the two halls, respectively. Thus, we can see that the heuristic parameter search and vanilla neural surrogate are inefficient to find the optimal configuration under the same computation time. Although the manual calibration reduces the MAE to 1.32\textdegree{}C to 1.1\textdegree{}C, it is labor-intensive. In addition, its MAEs are higher than Kalibre's. The lower MAEs achieved by Kalibre further improves the fidelity of the CFD results. If such results are used to guide data center operations, the risks caused by the errors can be further reduced. In sum, systematic approaches to improve the fidelity of data center digital twin are always desirable.

\subsubsection{Compute time}

From the results in Table~\ref{tab4}, Kalibre achieves sub-1\textdegree{}C MAEs with 10 calibration iterations and the lowest MAEs with about 15 calibration iterations. The compute time breakdown of each iteration is as follows: about 200 seconds for surrogate training, about 100 seconds for configuration search, about 26 minutes for CFD model solving with 32 CPU cores. Kalibre's compute time for calibrating a hall's CFD is about 5 hours and 7.5 hours for 10 and 15 iterations, respectively. For vanilla neural net to reach similar accuracy, it requires more iterations, resulting in 3x$\sim$5x compute time compared with Kalibre's to generate enough data for training.

\section{Discussions and Conclusion}
\label{sec:discuss}

\begin{table}[t]
	\caption{MAE achieved with 10 calibration iterations, as well as the lowest MAE achieved and the needed iterations.}
	\label{tab4}
	\begin{tabular}{cccccc}
		\toprule
		\multirow{2}{*}{} & \multirow{2}{*}{Approach}  & MAE (\textdegree{}C) & MAE (\textdegree{}C) & Needed  & \multirow{2}{*}{Auto?} \\
		& &  10 iters & lowest & iters \\
		\midrule
		\multirow{5}{*}{Hall A} & Manual & 1.32 & N/A & N/A & \ding{53} \\
		& Heuristic & 4.75 & 3.49 & 127 & \checkmark \\
		& Vanilla & 1.44 & 1.09 & 50 & \checkmark \\
		& \textbf{Kalibre} & \textbf{0.81} & \textbf{0.76} & \textbf{14} & \checkmark \\
		\midrule
		\multirow{5}{*}{Hall B} & Manual & 1.1 & N/A & N/A & \ding{53} \\
		& Heuristic & 3.80 & 2.61 & 130 & \checkmark \\
		& Vanilla & 1.33  & 0.80 & 28 & \checkmark \\
		& \textbf{Kalibre} & \textbf{0.75} & \textbf{0.59} & \textbf{15} & \checkmark \\
		\bottomrule
	\end{tabular}
\end{table}

We now discuss several worth-noting issues.
First, the primary purpose of the surrogate is to improve the efficiency of parameter search. The surrogate does not provide a full-fledged approximation of the CFD model. For instance, it does not model the temperatures at the locations without sensors, which are modeled by the CFD model in contrast. Thus, only the calibrated CFD model shall be used as a digital twin for the run-time temperature evaluation in improving energy efficiency and reducing operational risks. 
Second, the surrogate architecture described in this paper is for data halls with hot-aisle containment. Thus, heat recirculation is not considered. To address the data halls without hot-aisle containment, heat recirculation and temperature mixing effects should be added to the neural surrogate's design.
Third, this paper mainly focuses on temperature prediction. For other types of prediction, Kalibre can be extended to address their calibration problems with proper surrogate designs. For instance, if air flow rate sensors are deployed, Kalibre can be extended to calibrate the CFD for predicting air velocity distribution.

In conclusion, this paper presents Kalibre, an automatic surrogate-based approach to calibrate data center CFD models. The design of Kalibre's neural surrogate integrates prior knowledge including the thermal relations among the key variables of the physical infrastructure. Thus, it reduces the demand on the amount of training data generated by the compute-intensive CFD model solving. We demonstrate its effectiveness on two CFD models built for two production data halls that host thousands of servers. The CFD models calibrated by Kalibre achieve temperature prediction MAEs of 0.81\textdegree{}C and 0.75\textdegree{}C, respectively. Compared with manual calibration, Kalibre's improvement of up to 0.5\textdegree{}C is significant in CFD modeling due to the sharply increased difficulty in improving accuracy when the errors are already low (i.e., at around 1\textdegree{}C). Kalibre sheds lights on the calibration of other compute-intensive models to pursue high accuracy in approximating complex physical processes.

\begin{acks}
	This work is funded by National Research Foundation (NRF) via the Green Data Centre Research (GDCR) and the Green Buildings Innovation Cluster (GBIC), administered by Info-communications Media Development Authority (IMDA) and Building and Construction Authority (BCA) respectively.
\end{acks}

\bibliographystyle{ACM-Reference-Format}
\bibliography{ref}

\end{document}